\documentstyle[aasms4]{article}

\begin{document}

\lefthead{Shear Turbulence in MBM16}
\righthead{LAROSA, SHORE, \& MAGNANI}

\title{A Dynamical Study of the Non-Star Forming Translucent Molecular
Cloud MBM16: Evidence for Shear Driven Turbulence in the Interstellar
Medium} 

\author{T. N. LaRosa\altaffilmark{1}, Steven N. Shore\altaffilmark{2},
and Loris Magnani\altaffilmark{3}}

\altaffiltext{1}{Department of Biological and Physical Sciences,
Kennesaw State University, 1000 Chastain Road, Kennesaw, GA 30144
(ted$@$avatar.kennesaw.edu)} 

\altaffiltext{2}{Department of Physics and Astronomy, Indiana
University South Bend, 1700 Mishawaka Avenue, South Bend, IN
46634-7111 (sshore$@$paladin.iusb.edu)} 

\altaffiltext{3}{Department of Physics and Astronomy, University of
Georgia, Athens, GA 30602 (loris$@$zeus.physast.uga.edu)}

\begin{abstract}

We present the results of a velocity correlation study of the high
latitude cloud MBM16 using a fully sampled $^{12}$CO map, supplemented
by new $^{13}$CO data.   We find a correlation length of 0.4 pc.  This
is similar in size to the formaldehyde clumps described in our
previous study.  We associate this correlated motion with coherent
structures within the turbulent flow.  Such structures are generated
by free shear flows.  Their presence in this non-star forming cloud
indicates that kinetic energy is being supplied to the internal
turbulence by an external shear flow.  Such large scale driving over
long times is a possible solution to the dissipation problem for
molecular cloud turbulence.

\end{abstract} 


\newpage
\section{Introduction}

Superthermal linewidths are commonly observed in molecular clouds,
whether self-gravitating or not. Both internal and external driving
mechanisms have been suggested as the source of these velocities (cf.
Norman \& Silk 1980; Fleck 1983; Henriksen \& Turner 1984; Falgarone
\& Puget 1986; Scalo 1987; Mouschovias 1987; Myers \& Goodman 1988;
Falgarone \& Phillips 1990; Elmegreen 1990; Passot et al. 1995).
Virtually every suggested mechanism for producing the observed
linewidths requires mass-motions and/or non-linear wave motions both
of which are dissipative (Field 1979; Zweibel \& Josafatsson 1983;
Ghosh et al. 1994). Recent simulations by MacLow et al. (1998) and
Padoan \& Nordlund (1998) find that MHD turbulence decays on shorter
timescales than previously thought.  Thus, interstellar turbulence
requires continuous injection of kinetic energy.  The most obvious
source in large clouds is internal star formation.  There is, however,
a type of molecular cloud, the high latitude translucent variety (van
Dishoeck et al. 1991), that show the same superthermal velocity fields
but do not display evidence of star forming activity (e.g. Magnani et
al. 1995). In the absence of internal driving mechanisms such as star
formation or gravity, the motion must be continually produced by an
{\it external} agent. 

It is for this reason that we have carried out a dynamical study of
the translucent molecular cloud MBM16 (Magnani, Blitz, \& Mundy 1985,
hereafter referred to as MBM).  Our choice of this cloud is quite
deliberate. In our earlier paper (Magnani, LaRosa, \& Shore 1993,
hereafter referred to as MLS93) we reported finding several
large-scale formaldehyde clumps in MBM16.  These observations in
$H_2CO$ indicate that the clumps have a higher density than the rest
of the cloud. We  concluded that MBM16 is not self-gravitating and
that its internal structures are transient. Each clump has a distinct
average centroid velocity that differs from the others by about 0.5 km
s$^{-1}$. The difference is not due to a large scale systematic
motion. To quantify this, we have examined the standard statistical
measures for the line profiles.  The velocity centroid dispersion
within each clump, $\sigma_{c}({\rm H_{2}CO})$ in the notation of
Kleiner and Dickman (1985, and references therein) was found to be
quite low, $<$0.2 km s$^{-1}$, relative to the cloud as a whole
determined from $^{12}$CO data, 0.44 km s$^{-1}$, and relative to the
clump to clump centroid velocity difference of 0.5 km/s.   The CO data
were, however, based on a severely undersampled map. In this study, we
present the results of a more extensive $^{12}$CO map of the cloud
supplemented by new $^{13}$CO observations of the densest formaldehyde
clump that support and extend our earlier conclusions. 

\section{Observations} 

A complete map of MBM16 was made in the $^{12}$CO (J=1-0) transition
using the 1.2 m Harvard-Smithsonian Center for Astrophysics Millimeter
Wave Telescope. The telescope front end configuration is described by
Dame (1995). The map is a 6$^\circ \times$ 6$^\circ$ grid in Galactic
longitude and latitude ($\ell =$ 168$^\circ$ to 174$^\circ$ and $b =$
$-$34$^\circ$ to $-$40$^\circ$) approximately centered on the cloud.
The beamwidth of the telescope at 115 GHz is 8.4 arcmin and the
separation between the samples was 3.6 arcmin for a total of 97
$\times$ 97 points or 9409 spectra.  At the assumed cloud distance of
80 pc (Hobbs et al. 1988) 8.4 arcmin corresponds to a linear
resolution of 0.2 pc. The data were taken in frequency-switched mode
at a velocity resolution of 0.65 km s$^{-1}$. The total bandwidth was
150 km s$^{-1}$ centered at 0 km s$^{-1}$ with respect to the LSR.
Typical rms noise values for each spectrum were in the 0.133 to 0.281
K range. 

A contour map of the integrated antenna temperature ($\int T_A^* \
dv$) is shown in Figure 1.  The velocity map is shown in Figure 2. MBM
16 is clearly centered in the map, and is completely enclosed by the
mapping pattern. The general shape of the cloud is similar to the
poorly-sampled map by MBM, and the ragged structure is reminiscent of
other high latitude clouds.  At the bottom left of the field in Figure
1 is a heretofore unidentified high latitude cloud. Since the LSR
velocity of the new cloud is significantly different from that of MBM
16, it is likely that the two clouds are at different distances. The
molecular gas comprising MBM 16 ranges in velocity from 4 to 10 km
s$^{-1}$, while the new cloud LSR velocities lie between -20 and -5 km
s$^{-1}$.  Similarly, in the northeast corner of Figure 1 is a
molecular feature which has LSR velocities in the 12-14 km s$^{-1}$
range. CO channel maps of this region indicate that this feature is
also likely to be a separate molecular cloud (Magnani et al. 1998, in
preparation). 

We have also mapped the largest formaldehyde clump and a portion of
another in $^{13}$CO at higher resolution than we reported in MLS93.
These data were obtained during two observing sessions at the
NRAO\footnote{The National Radio Astronomy Observatory is operated by
Associated Universities, Inc. in a cooperative agreement with the
National Science Foundation.} 12 meter telescope at Kitt Peak,
Arizona, 1992 Dec. 26 - 1993 Jan. 1 and 1993 July 2 - 7.  The 3 mm
receiver with an SIS mixer operated with a total system plus sky
temperature between 300 and 700 K.  Dual polarizations were fed into a
100 kHz filter back end, providing a velocity resolution of 0.26 km
s$^{-1}$.  The data were obtained in frequency switched mode with
separation of $\pm$2 MHz.  The polarizations were averaged together
and the resulting spectrum was folded resulting in typical {\it rms}
noise temperatures from 10 to 30 mK after Hann smoothing. We used a
uniformly spaced 20$\times$27 grid with 3 arcmin separation between
beam centers (the beam size is $\sim$ 1 arcmin at 113 GHz).  The map
concentrated primarily on clump 3 in the H$_{2}$CO study described in
MLS93, although a portion of clump 2 was included in the western
section of the map.   The integrated $^{13}$CO(J=1-0) intensity map of
the region is shown in Figure 3. Line emission was detected in about
200 of the 540 map positions.

\section{Statistical and Correlation Analyses} 

For fully developed turbulence, over a large enough region the
dispersion in the centroid velocities should be about the same as the
individual line widths.  The $^{13}$CO data agrees with the H$_{2}$CO
results in showing a much smaller centroid dispersion within the
clumps. Following Kleiner \& Dickman (1985), for the entire sampled
cloud the parent velocity dispersion of the mapped region, $\sigma_p$
({\it i.e.}, the average dispersion of the individual spectra,
$\sigma_i$, is 0.25 km s$^{-1}$, and the dispersion in the intensity
weighted velocity centroids determined from Gaussian profile fitting,
$\sigma_c$, is 0.31 km s$^{-1}$.  For the  leftmost clump located at
$\Delta$RA $\ge$ 50.25) in Fig. 3, $\sigma_{i}$ = 0.21 and 
$\sigma_{c}$ = 0.10 km s$^{-1}$. The implication is that the clumps
are coherent in velocity.  The $^{13}$CO map is, however, too small to
be useful for a correlation study.  Our new $^{12}$CO map is fully
sampled and sufficiently large to permit such an analysis.

The velocity correlation function is the simplest description of the
dynamical properties of a flow.  By definition, the spatial velocity
correlation function for a homogeneous turbulent flow is:
\begin{equation} 
c_{ii}(\tau ) = <[v_{i}(r)-<v_{i}(r)>][v_{i}(r+\tau ) - <v_{i}(r)>]> 
\end{equation} 
where the angle brackets denote the
spatial averages.  The lag, $\tau$, is a scalar quantity.  We are, of
course, limited to measuring only the radial velocities so we can
construct only the longitudinal velocity autocorrelation along the
line of sight, $c_{zz}(\tau )$. We normalize the autocorrelation
function by its value at zero lag, $C(\tau ) = c_{zz}(\tau
)/c_{zz}(0)$. In standard terminology, this is known as the biased
autocorrelation function (see Miesch \& Bally 1994, hereafter referred
to as MB94). 

Before performing the autcorrelation analysis, we trimmed  the data to
remove points in the map with velocities that are clearly in excess of
the bulk of the cloud.  In this case, we retained all points with a
signal to noise ratio of 4.5$\sigma$ or higher with $4.0 \le v_{rad}
\le 10.0$ km s$^{-1}$, the range of velocities found within MBM16. 

As discussed in Scalo (1984) and further elaborated in MB94, any
correlation analysis requires some filtering of the dataset to remove
the large scale trends.  Therefore, we have applied two different
methods to the MBM16 data.  In the first procedure, we subtracted the
mean velocity, 7.24 km s$^{-1}$, from the trimmed map to form the
fluctuation map. In the second procedure we subtracted a smoothed map
from the data that was obtained from the trimmed map by taking a
running mean filter of preset set size that included only detected
points.  This is the procedure used by Miesch and Bally. We then
carried out the autocorrelation analysis of these filtered $^{12}$CO
data using two independent methods.  The first, which is identical to
the one described in MB94, used the two dimensional fast Fourier
transform (FFT) in IDL.  For comparison, in the second we used the
scalar algebraic correlation function. 

The results are plotted in Figures 4 and 5.  Figure 4{\it a-c} shows
the two dimensional correlation results for (a) the mean subtracted,
(b) the 15 point smoothed, and (c) the 30 point smoothed maps.  In
view of the agreement between the different procedures, and the lack
of systematic trends in the velocity, we use method (a) for Figure 5
and the subsequent analysis.  In Figure 5, we show the comparison
between the scalar (1D) and 2D correlation funnctions for the mean
subtracted map.  The two methods agree.  There is a detectable
correlation up to 5 lags or 18.0 arcmin, which corresponds to a
physical length of 0.42 pc\footnote{This is about 2.5 beamwidths.  We
used the same random maps described below to test the effects of beam
size on our derived correlation functions. For example, for a gaussian
beamwidth (FWHM) of 4 lags convolved with the random map, the
correlation function has its first zero at 2 lags.  The same map
convolved with our beam size yielded no correlation above 1 lag.
Therefore, tests show that the correlation we are reporting here does
not arise as an artifact of the sampling.}. To further understand what
the correlation function is measuring, and in particular how to define
the correlation length, we performed a series of simulations for which
the results are shown in Figure 6.   To generate a test random map
with the same spatial distribution as the real data, we took the
trimmed map, normalized it to unity for all detections, and then
multiplied each detected location by a gaussian random number with a
unit dispersion.  The correlation function for this completely random
map is shown in Fig. 6a.  Notice that although the velocity is
spatially confined to the regions where CO was detected, there is no
hint of any spatial stucture in the Figure. We then simulated the
presence of a coherent structure within this map by setting all
velocities in a 10$\times$10 region equal to a constant and, in a
series of trials, varied the constant.  This creates a coherently
moving clump of known size and kinetic energy.  The velocity was
increased from 0 to 2 km s$^{-1}$ to produce a set of maps that were
then run through the same analysis as the real data.  What they show
is that there are clear limits to the information that can be derived
from a correlation analysis, but that the detection of a correlated
flow is possible. 

For instance, in Figures 6{\it b-d}, the amplitude of the velocity was
increased from 0.5 km s$^{-1}$ to 2.0 km s$^{-1}$ where the random map
has velocities in the range -6.3 to 4.4 km s$^{-1}$.  Again, we note
that the velocity dispersion was 1 km s$^{-1}$.  With an amplitude of
0.5 km s$^{-1}$, the clump had about 3\% of the total power in the
map, and the structure was clearly undetectable (Fig. 6b). For 1 km
s$^{-1}$, or about 13\% of the total power, the structure is obvious
(Fig. 6c), and becomes more so with increasing energy.  An additional
result of the simulations is that the size of the coherent region is
not the $e^{-1}$ point but where the correlation function has its
first zero (e.g. Townsend 1976).   More important is the result that
the lack of detection of a coherent region by the correlation method
merely places an upper limit on the size and power of such regions,
but does not rule out their presence. 

\section{Discussion}

On the basis of a purely statistical analysis of the line profiles 
for a number of molecular species, $^{12}$CO, $^{13}$CO, and H$_{2}$CO, 
it is clear that MBM16 contains a number of regions with exceptionally 
low dispersions in the line centroid velocity.  We emphasize that these 
are not simply ``quiet spots'' in the flow.  The individual profiles 
throughout these regions have the same widths as the rest of the cloud, 
which is much larger than the dispersion in the centroids.  The implication 
of this observation is that the size-line width relation (Larson 1981) 
breaks down for 
this cloud.  This is shown in detail in Figure 7.  There is no systematic 
trend to the data, and those regions identified in MLS93 from the 
formaldehyde observations clearly stand out in the  $^{12}$CO data 
as anomalously underdispersed 
for their size.

The results of our correlation analysis support and strengthen these 
statistical conclusions.  We find a correlation length of 
roughly 18 arcmin, corresponding to a physical scale of 
about 0.4 pc.  Our simulations suggest that 
these regions must together contain at least 10\% of the kinetic 
energy of the cloud.   We argue that these regions are the 
interstellar analogs of the {\it coherent structures} that 
usually accompany turbulence generated in free and bounded shear 
flows.  These have been the subject of increasing attention in 
the past few years as their importance in turbulence 
has been more clearly recognized in terrestrial flows 
(e.g. Lumley 1989; George \& Arendt 1991; 
Robinson 1991; McComb 1991; 
Shore 1992; Holmes, Lumley, \& Berkooz 1996).  

In the most general case where the flow can be driven both internally 
and externally, it is likely that no single source scale would be 
detected using correlation methods because none exists.  Instead, 
the variety of processes act on so many different scales that the 
no one produces a clean dynamical signature.  As far 
we know, there have been 
no previous dynamical studies of any {\it isolated} non-star forming 
cloud\footnote{Although Heiles 2 and other non-star forming clouds have 
been observed, these are all associated with or imbedded within 
larger star forming cloud complexes.   MBM16 is completely isolated.}.  
It is therefore not surprizing that no unambiguous velocity 
correlation length is 
generally found (e.g. MB94).  Scalo (1984) 
suggested that a length scale of order 0.3 pc in the $\rho$ Oph cloud 
might be due to turbulence, although it was possible that large scale 
structure could produce a similar signature.  It is precisely 
the simplicity of the velocity field in MBM16 that makes such a 
detection more likely if it exists.  
This cloud is nearby so we can obtain fully sampled maps of 
comparatively small scale structures.  More important, the 
lack of internal sources for turbulence in MBM16 means that any 
external driver will leave a recognizable signature in the velocity 
field of the cloud.  

If we are correct in assigning the source of the coherent 
structures to an 
external shear flow, then there are important consequences for the 
interpretation of the line profiles in this and other clouds.  First, 
such phenomena should be ubiquitous in the interstellar medium.  For 
instance, 
any H I cloud should show similar structures.  Boundary 
layer effects have already been invoked in chemical studies of 
molecular clouds (e.g. Charnley et al. 1990).  In 
particular, we suggest that a correlation analysis of a high velocity 
H I cloud might reveal coherent structures generated by boundary 
shear with the interstellar medium\footnote{Recent analytic and numerical 
work by Vietri, Ferrara, \& Miniati (1997) suggests that non-self 
gravitating clouds are not disrupted by the Kelvin-Helmholtz instability.}.  
Second, such driven turbulence will 
produce MHD waves so long as the fluid and magnetic field are 
coupled.  This is true throughout the interstellar medium.  Third, 
 the fact that 
large scale external driving can feed the observed 
internal motions points to a long-lived 
source.  Even though both MHD waves and turbulence are intrinsically 
dissipative, the longevity of the driver can overcome the 
dissipation and maintain the stability of the clouds for very long times.  
Thus, the detection of turbulence does not imply any 
particular youth for the clouds.

\section{Summary}

Our purpose in this paper has been to characterize the velocity field
within a non-star forming molecular cloud.
We were guided by earlier H$_2$CO
observations at 6 arcmin resolution (0.14 pc) that showed dynamical
ordering on a length scale of about 0.5 pc.  
Using higher resolution and a far more complete $^{12}$CO map, we find 
evidence for a velocity correlation on a scale of 0.4 pc.  We attribute 
this to the formation of coherent structures in a externally driven 
turbulent shear flow.  This could easily be 
 the source for both MHD and fluid turbulence 
on small scale within this and other molecular 
clouds\footnote{This explanation may specifically apply 
to giant molecular clouds.  For
instance, Williams, Blitz, \& Stark (1995) have observed numerous 
clumped structures in turbulent clouds that do not contain embedded 
stars yet otherwise display the same turbulence properties as we have 
discussed here.  We thank the referee, T. Harquist, for suggesting this 
addition.} and supply kinetic 
energy to the clouds on potentially long timescales.

\acknowledgments

We thank Paul Hart, Tom Fokkers and Dale Emerson, without whose expert
help the observations would not have happened.  We also thank John
Scalo for extensive and selfless advice and correspondence, and Tom 
Harquist, the referee, for his advice.  We also thank Ellen
Zweibel, Alyssa Goodman, Enrique V\'azquez-Semadeni, Bruce Elmegreen,
Jose Franco, and Michael P\'erault for discussions.  We thank Dap
Hartmann for help with the figures of the $^{12}$CO observations. TNL
is supported by a NASA JOVE grant to Kennesaw State University.  SNS
is partially funded by NASA. 

\newpage 

\clearpage

\section*{Figure captions}

Fig.\ 1. $^{12}$CO J=1-0 contour map of a 6$^{o}$$\times$6$^{o}$
region centered on MBM16.  The velocity range of the map is $-10  \le
v_{\rm LSR} \le 15$ km s$^{-1}$.  The contours are at [10, 30, 50, 70,
80, 90, 95] percent of the maximum CO intensity of 7.23 K km s$^{-1}$.
 The objects to the extreme northeast and southeast of the map are
probably not associated with MBM16 (see text).  The size of the 8.4
armin beam, the effective map resolution, is shown in the upper right
corner. \\ 

Fig.\ 2.  $^{12}$CO velocity map corresponding to the cloud shown in
Fig. 1.  This map corresponds to the velocity range 4 to 10 km
s$^{-1}$.  The orientation is the same as Fig. 1.  One lag corresponds
to 3.6 arcmin for the fully sampled map.\\ 

Fig.\ 3.  The $^{13}$CO(J=1-0) velocity map of a subregion in MBM16.
The (0,0) location corresponds to  ($\alpha$,$\delta$) $=$ (3$^h$
21$^m$ 48$^s$, 10$^\circ$ 42' 00")  in 1950 coordinates. The region
covered by the map encompasses all of clump 3 and the eastern portion
of clump 2 as defined in MLS93. The sampling is every 3' with a 1'
beam and surface plot shows all velocities greater than 5 km s$^{-1}$
(maximum is 8 km s$^{-1}$).  \\ 

Fig.\ 4.  Two dimensional velocity correlation function for different
normalizations (see text) obtained by (a) subtracting a constant mean
for the cloud of 7.24 km s$^{-1}$; (b) subtracting a smoothed map with
a 15 point running mean; and (c) subtracting a 30 point running
mean.\\ 

Fig.\ 5. Comparison of the one and two dimensional velocity
correlation functions for the mean subtracted data.  Note that the
scalar correlation function, the solid line in the figure, is {\it
not} a fit to the points.\\ 

Fig.\ 6. Results of simulations of a coherent structure in a normally
distributed velocity fluctuation map.  The figure shows the 2D
velocity correlation function for simulations described in the text
with (a) $\Delta v$ = 0 km s$^{-1}$; (b) 0.5 km s$^{-1}$; 1.0 km
s$^{-1}$; (d) 2 km s$^{-1}$.\\

Fig.\ 7.  Size-line width results for the $^{12}$CO data for MBM16.
The squares represent the H$_{2}$CO clumps while the crosses are for
regions throughout the cloud.  The length scale is the square root of
the number of detected points within the sampled region.  The The
centroid velocity dispersion was determined from the intensity
weighted centroids of the individual line profiles.  Notice the
distinct separation of the clumps from the bulk of the data and the
absence of any obvious overall trend.\\

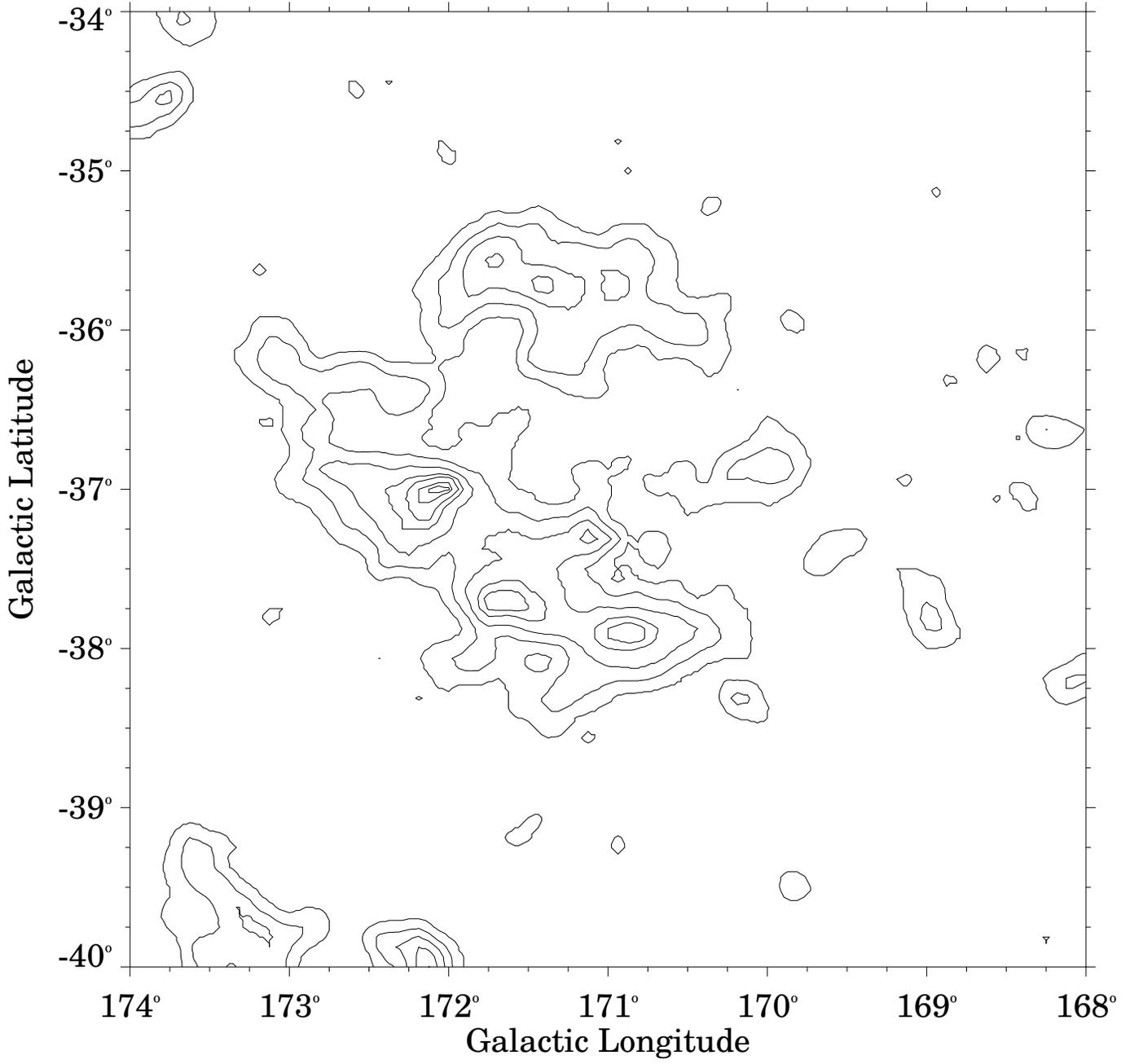
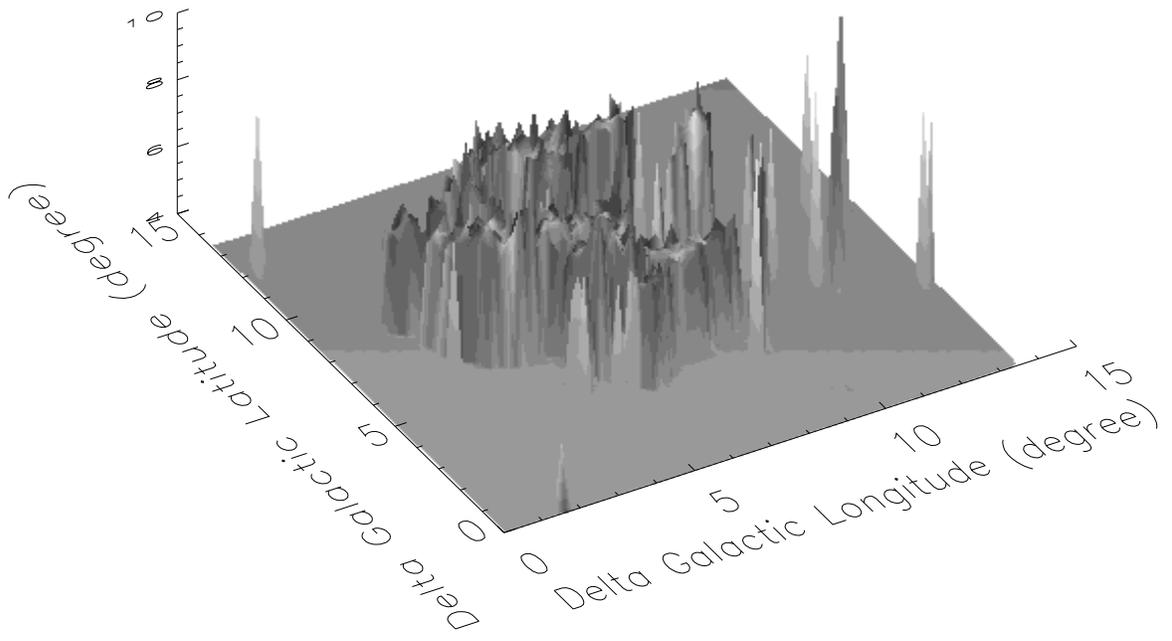
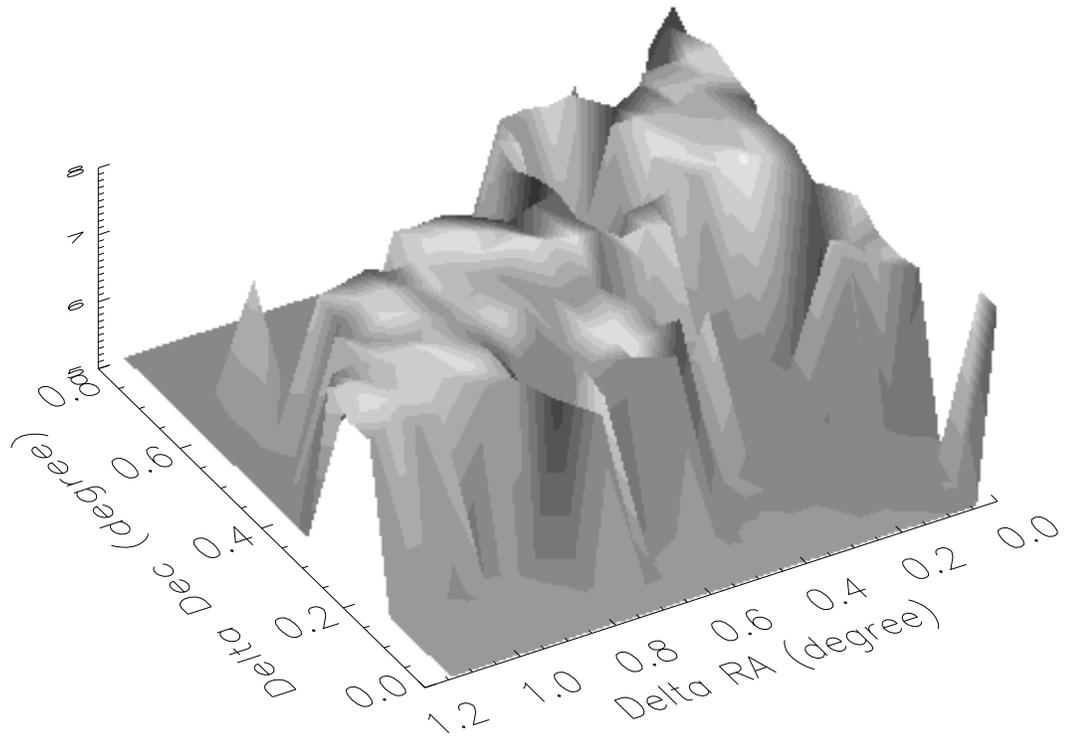
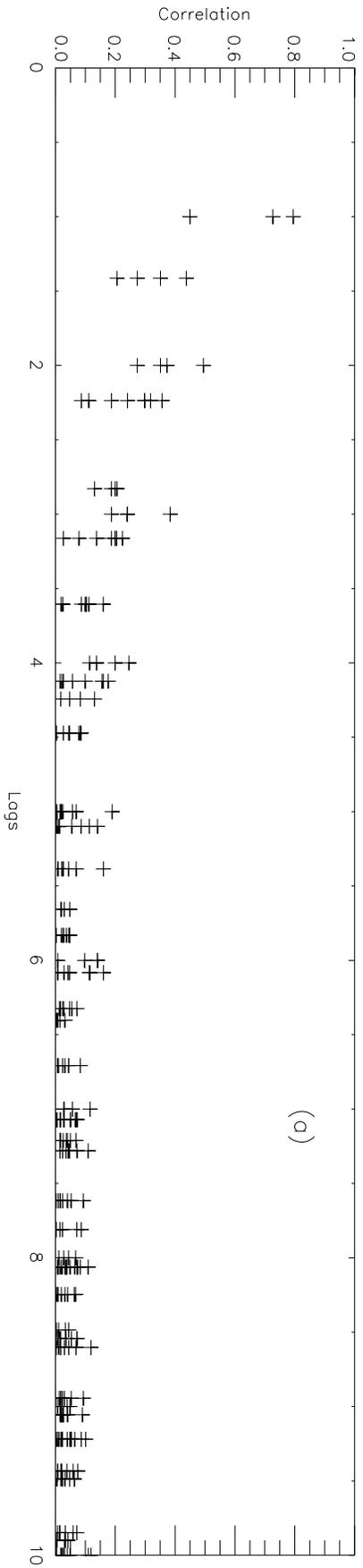
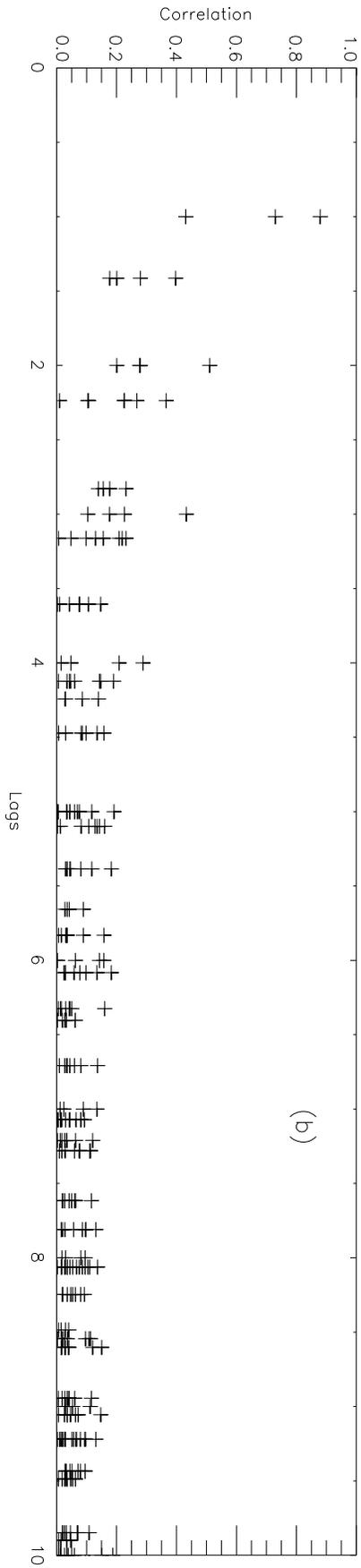
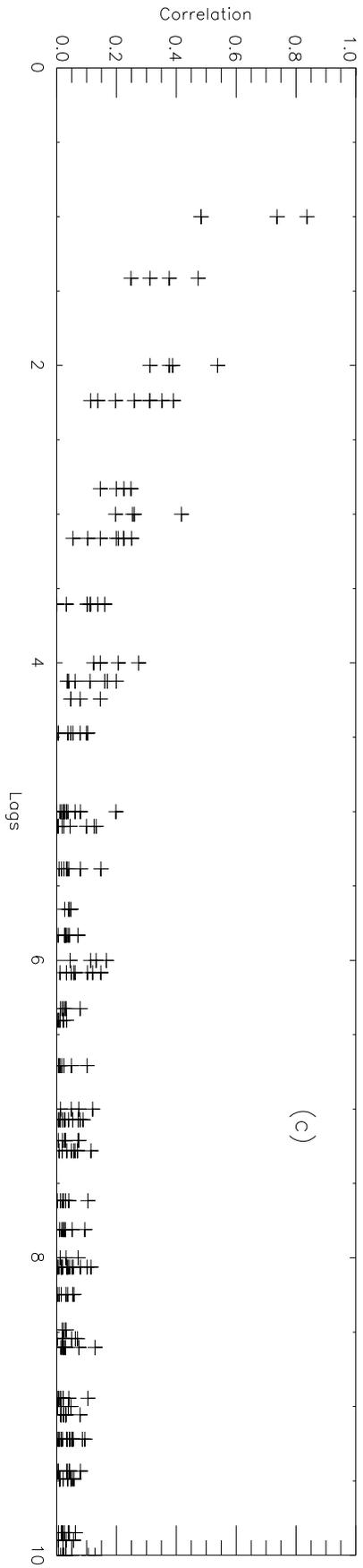
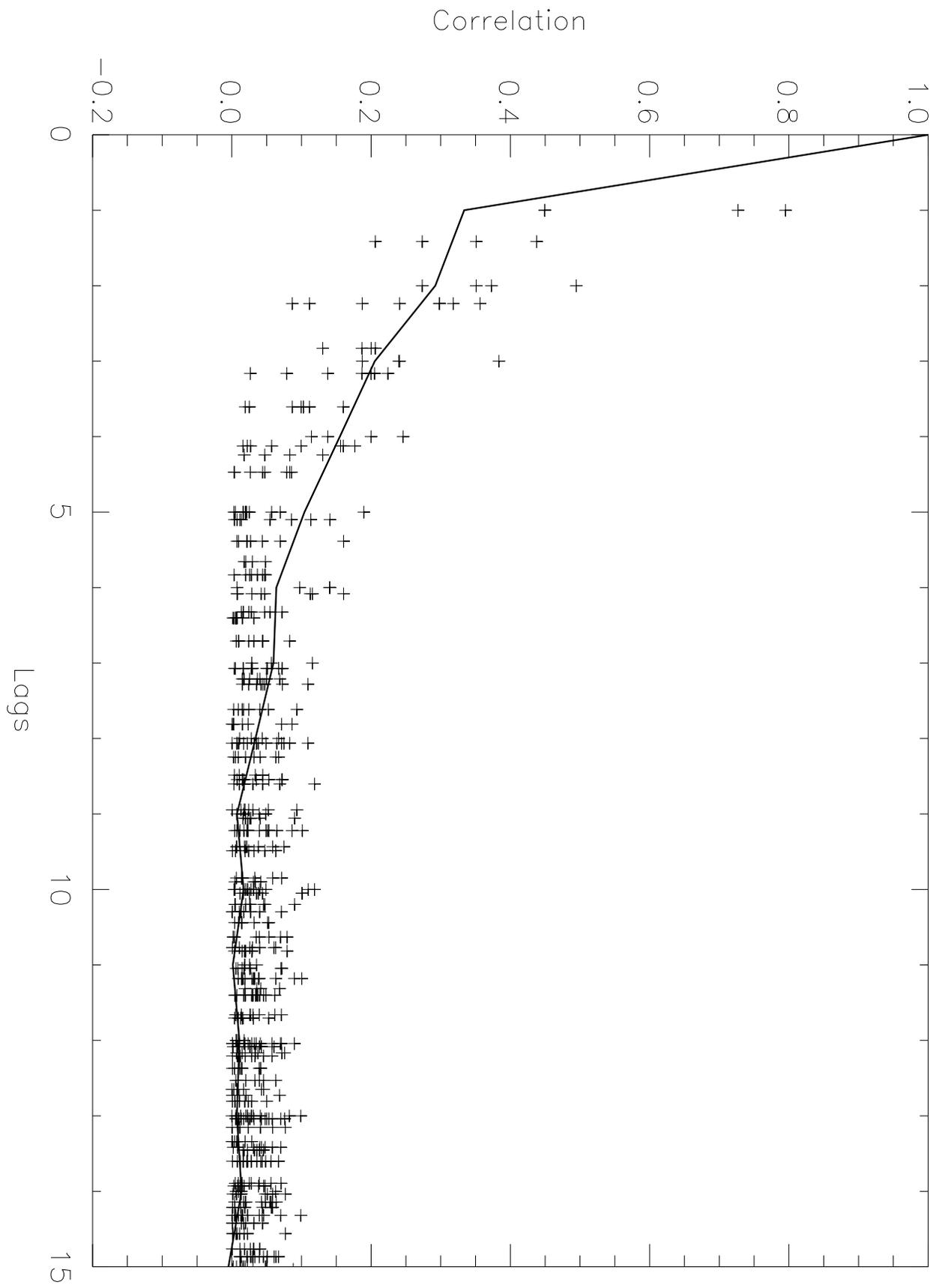
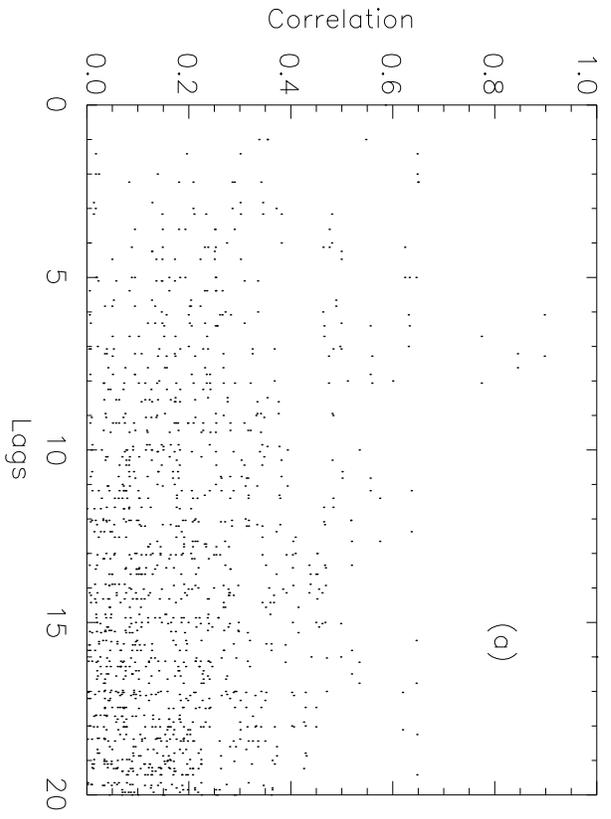
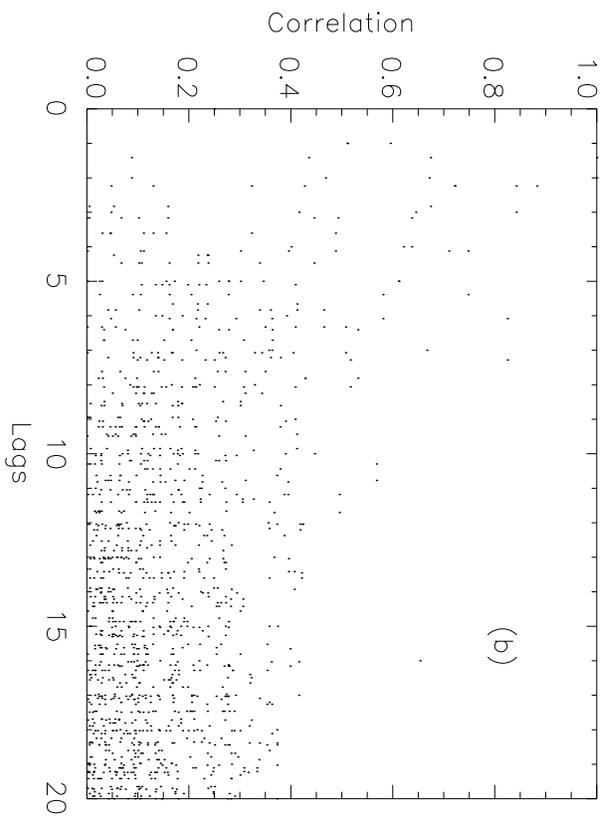
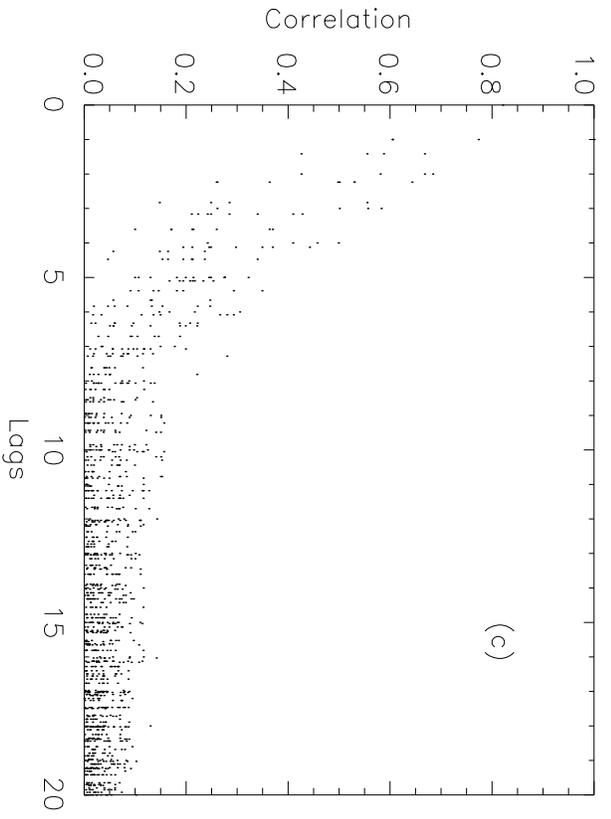
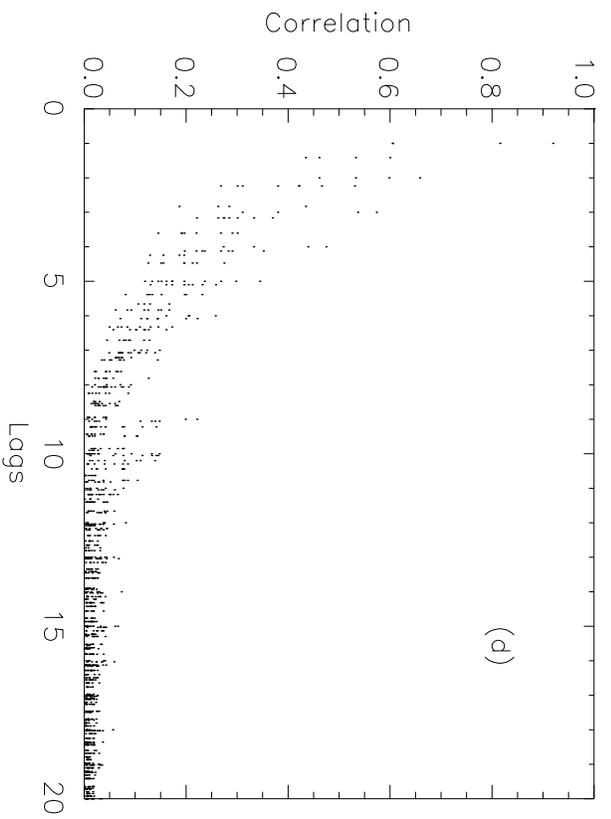
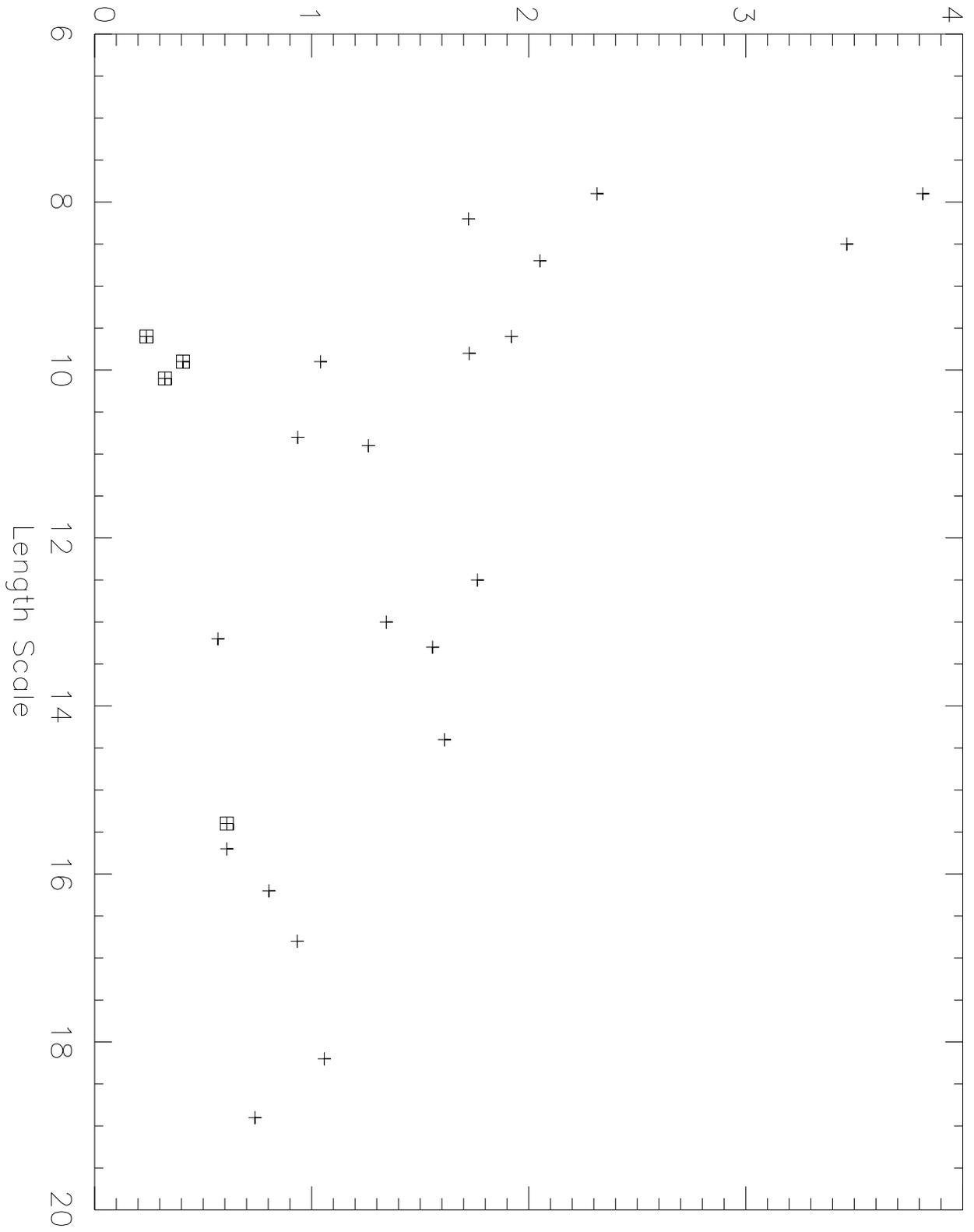
\end{document}